\documentclass[a4paper,11pt]{article}
\usepackage{pos}
\usepackage{enumitem}
\title{The Very-high-energy Open Data Format: towards a shared, open data format in very-high-energy astronomy}
 \ShortTitle{The Very-high-energy Open Data Format}

\author*[a,1]{B. Khélifi}
\author[b]{R. Zanin}
\author[c,b]{K. Kosack}
\author[d]{L. Olivera-Nieto}
\author[e]{J. Schnabel}

\affiliation[a]{Université Paris Cité, 
CNRS, Astroparticule et Cosmologie, F-75013 Paris, France}

\affiliation[b]{Cherenkov Telescope Array Observatory gGmbH, 
Via Gobetti, Bologna, Italy}

\affiliation[c]{IRFU, 
CEA, Université Paris-Saclay, F-91191 Gif-sur-Yvette, France}

\affiliation[d]{Max-Planck-Institut für Kernphysik, 
P.O. Box 103980, D 69029 Heidelberg, Germany}

\affiliation[e]{ECAP, 
Friedrich-Alexander-Universität Erlangen-Nürnberg, Nikolaus-Fiebiger-Str. 2 , 91058 Erlangen, Germany}

\note{~For the VODF Steering Committee}

\emailAdd{khelifi@in2p3.fr}

\abstract{In very-high-energy (VHE) gamma-ray astronomy, the community is converging towards the use of a common open data format, called "Data formats for Gamma-ray Astronomy", for the high-level data products. This format is in use for ground-based TeV observatories like H.E.S.S, MAGIC or HAWC, some of whom plan to openly release high-level data products. These efforts are parallel to the development and use of open analysis software such as the Gammapy package. This open initiative has shown that it is possible to define common standards even without governance. With the advent of open VHE observatories (e.g. CTAO, KM3NeT) and an increase in both multi-wavelength and multi-messenger studies, such standards should evolve to support all of VHE multi-messenger astrophysics. For these reasons, a new initiative has been created to specify formats of high-level data from very and ultra high energy gamma-ray facilities and from VHE neutrino detectors. It also aims to better respect the FAIR principles and the IVOA recommendations.This communication will present the Very-high-energy Open Data Format (VODF) project that has been established by eleven VHE astroparticle facilities. Its structure, its organisation and its goal will be presented. Anchored in Open Science, our goal is to solicit comments and future contributions from the VHE astrophysics community.}

\ConferenceLogo{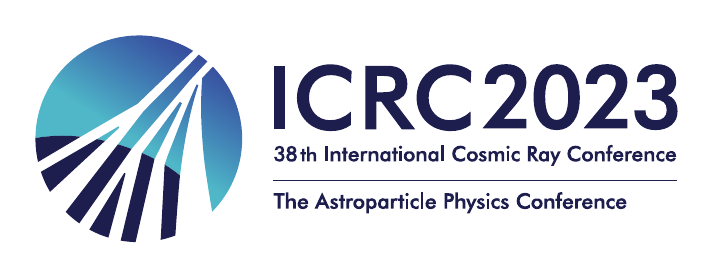}

\FullConference{%
38th International Cosmic Ray Conference (ICRC2023)\\
  26 July - 3 August, 2023\\
  Nagoya, Japan}

\begin{document}
\maketitle

\section{Landscape of the VHE data analysis}
The field of ground-based very-high-energy (VHE) gamma-ray astronomy is rapidly evolving with more than 300 detected sources~\citep{tevcat, lhaasocatalog} and with the commissioning and operation of large detector arrays, such as CTAO~\citep{cta} or LHAASO~\citep{lhaaso}. Known gamma-ray sources are associated with objects such as supernova remnants, pulsar wind nebulae, binary systems, novae, stellar clusters, star-forming regions, star-forming galaxies, active galactic nuclei, or gamma-ray bursts. Their studies often require the use of data from other wavelengths. In parallel, VHE neutrino astronomy has recently reached a stage that allows the study of neutrino emission from Galactic and Extragalactic sources~\citep{antares_gal, icecube_gp, icecube_extragal}, leading to multi-messenger astrophysical analyses of both isolated sources and large-scale emission. The next generation of more sensitive instruments~\citep{IceCube-Gen2, km3net-arca} is currently in deployment.

The improvement of sensitivities of VHE astrophysical facilities is accompanied by an increasing amount of public data (e.g.~\citep{hess_dr, antares_dr}) and open analysis libraries (e.g.~\citep{gammapy, 3ml}). The opening of research products (publications, data, software) is stimulated by national and transnational roadmaps for Open Science (e.g.~\citep{eu_os, nasa_os}) that recommend or require free access. Some newly created VHE observatories follow these requirements and will openly grant access to data and analysis software. These policies will stimulate even more the multi-wavelength and multi-messenger analyses. As demonstrated in \citep{hawccrab} with VHE gamma-ray data and in \citep{tim} using gamma-ray and neutrino data, both with the library Gammapy~\citep{doigammapy}, joint analysis of multi-instrument data can be performed with a coherent statistical treatment while dealing with systematics of/between measurements.
 
Such multi-instrument analyses are becoming a reality thanks to the use of a common open FITS-based\footnote{~\href{https://fits.gsfc.nasa.gov/fits\_home.html}{https://fits.gsfc.nasa.gov/fits\_home.html}}~\citep{fits} data format for the high level or science-ready data (see section~\ref{sec:dl}) of the VHE gamma-ray instruments, called "Data formats for Gamma-ray Astronomy" (or Gamma-ray Open Data Format, GADF)\footnote{~\href{https://gamma-astro-data-formats.readthedocs.io/en/v0.3/}{https://gamma-astro-data-formats.readthedocs.io/en/v0.3/}}. The work of data formatting started back in 2010 by TeV experts from H.E.S.S. and CTA following the example of the Fermi-LAT and OGIP formats. And it leads to the first version of GADF in 2016~\citep{gadf}. Today, the running gamma-ray experiments H.E.S.S., MAGIC, VERITAS and HAWC are using this data format for some of their data sets. The use of this data format for testing purposes and for scientific publications has shown that its design describes efficiently the VHE gamma-ray data and permits accurate scientific analyses. 
After few years of usage of this format, some main user and core contributors (GADF, CTAO, H.E.S.S., etc) realise the need to evolve because the GADF open initiative started to show some limitations~\citep{gadflimits}, both on the format requirements and its organisation.

Enriched by the gained experience of the GADF initiative, their main contributors contacted the current and under-deployment VHE gamma-ray and neutrino facilities to establish a new open initiative to build together an extended open data format for the VHE astroparticle experiments, the \emph{Very-high-energy Open Data Format} (VODF). Now officially supported by eleven VHE facilities, ASTRI, CTAO, FACT, Fermi-LAT, HAWC, H.E.S.S., IceCube, KM3NeT, MAGIC, SWGO and VERITAS, this newly born initiative has started its activities. This contribution describes its objectives and some of its requirements (section~\ref{sec:obj}), the targeted level of data that are considered to be formatted (section~\ref{sec:dl}), the organisation of the initiative (section~\ref{sec:org}) and the anticipated perspectives (section~\ref{sec:pers}).

\section{Guidelines of the open VODF initiative}
\label{sec:obj}

VODF is an open data model and format for VHE gamma-ray and neutrino astronomy. These experiments strongly differ by their detection techniques, their instrumentation and their type of raw data. Ground-based gamma-ray detectors fall into two categories, Imaging Atmospheric Cherenkov Telescopes (IACTs) and Water Cherenkov Detectors (WCDs), whereas neutrino detectors are using optical modules immersed in a km3-scale volume of ice or water. They all share the following high-level properties, after a low-level data processing consisting of calibration, reconstruction and background reduction: 
\begin{itemize}[noitemsep,nolistsep]
\item they provide a list of gamma-ray or neutrino candidates characterised by their arrival time, their energy and their celestial arrival direction,
\item their instrumental response to events can be characterised by the same quantities and have the same type of observational dependencies, such that their associated Instrument Response Functions (IRFs) can be factorised in the same manner,
\item as these instruments study the same astrophysical processes, the scientific products are similar: sky maps, spectra, light curves, along with associated statistical information like significance or likelihood profiles.
\end{itemize}
In this context, the formats can be homogeneous and shared. This approach has been tested and has shown its success to generate scientific results and in addition to offer real interoperability between instruments~\citep{gammapycrab, hawccrab}.

The goal of the VODF initiative is to provide a standard set of data models and file formats, starting at the reconstructed event level (\emph{science-ready data}) as well as higher-level products such as N-dimensional binned data cubes (including sky images, light curves, and spectra) and source catalogues. With these standards, common science tools can be used to analyse data from multiple high-energy instruments. \\

In parallel, several facilities will publicly open their data (e.g. CTAO, KM3NeT) or their archive~\citep{magicopen} and the use of certified repositories is strongly recommended. As a consequence, the data should be correctly curated such that they respect as closely as possible the FAIR principles~\citep{fair} (\emph{Findable, Accessible, Interoperable and Reusable}). Metadata describing the associated data are mandatory. VODF aims to standardise the format of these metadata for each level of data. 

In the field of astronomy, the community started to think about common standards in order to share and use more efficiently data. Formed in 2002, the \emph{International Virtual Observatory Alliance} (IVOA)\footnote{~\href{https://www.ivoa.net/}{https://www.ivoa.net/}} has the mission to create and maintain standards to "facilitate the international coordination and collaboration necessary for the development and deployment of the tools, systems and organizational structures necessary to enable the international utilization of astronomical archives as an integrated and interoperating virtual observatory". IVOA has established many advanced data models and standards that are an excellent guideline for our VHE domain (e.g. \citep{servillat2022provenance}). In addition, they offer new possibilities to find and access data with the Virtual Observatory (VO) services (such as Aladin or Topcat). 

Technical tests have been realised to respect more closely the FAIR principles and to use some of the IVOA standards. The H.E.S.S. test data release~\citep{hess_dr} or the ANTARES data release~\citep{antares_dr}  illustrate the new possibilities~\citep{servillat2022fair, juttavo} available to the new VHE observatories using standards set by astronomers.\\

Given this context of state-of-art data formating, the open initiative VODF is built on the following guidelines:
\begin{itemize}[noitemsep,nolistsep]
\item building a new VHE open data format common to IACT, WCD and neutrino instruments,
\item being compliant with the FAIR principles,
\item following as much as possible the IVOA standards,
\item working by consensus with open contributions under the supervision of supporting experiments.
\end{itemize}

\section{Data Levels}
\label{sec:dl}
\begin{figure}[h!]
\centering
\includegraphics[width=16cm]{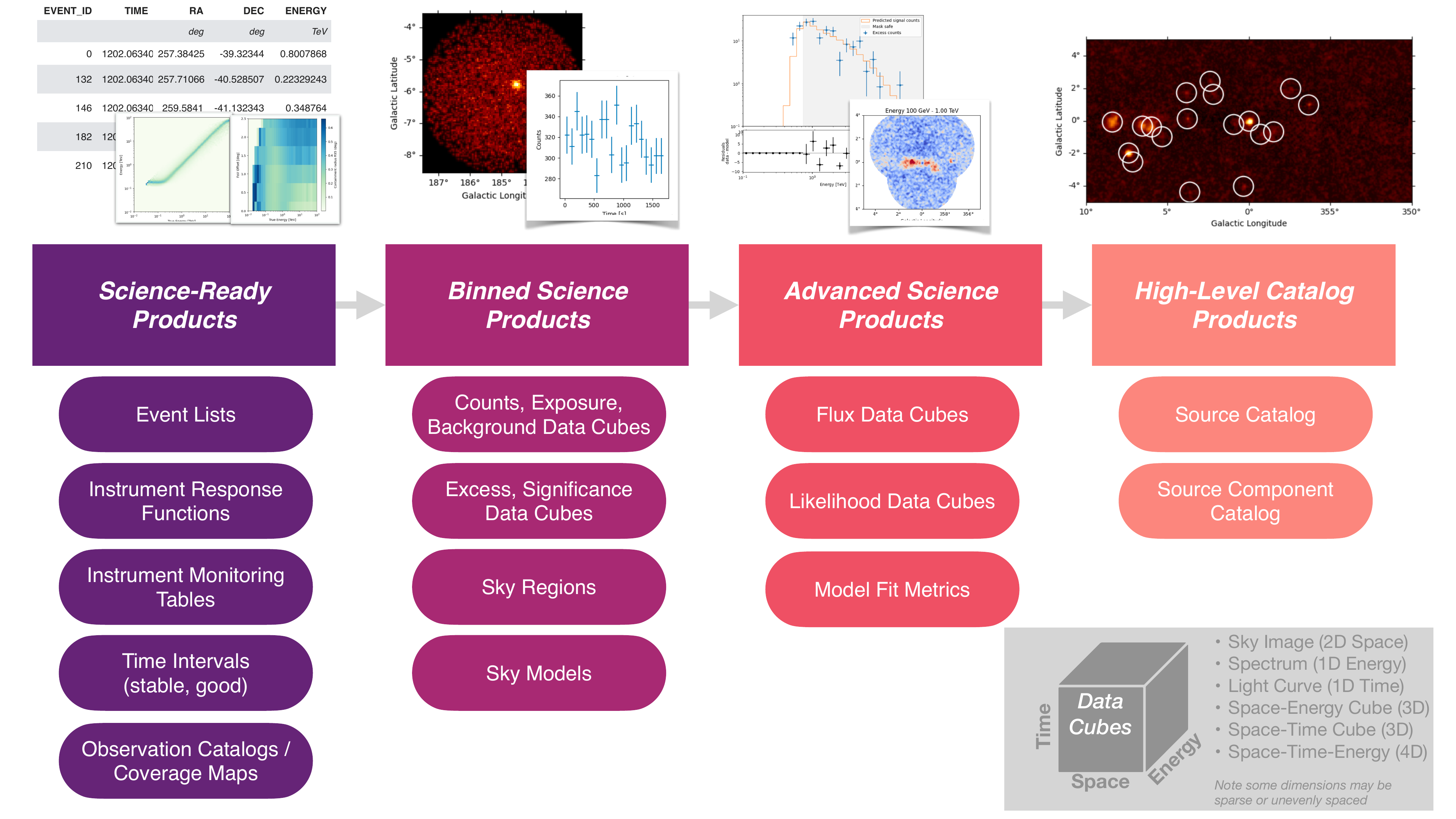}
\caption{Diagram of the different data levels. Each rounded box stands for data and their associated metadata.}
\label{fig:dl}
\end{figure}

VODF aims to settle standards of different levels of VHE data resulting from the calibration, reconstruction and background reduction of the instrument analysis pipelines. The data types and data levels to format are illusted in Fig.~\ref{fig:dl} and described below.

\begin{description}
\item[Common Data Structures.] The considered data are sharing common elements and structures. One can list in particular:
\begin{itemize}[noitemsep,nolistsep] 
\item the time formats, following the FITS standards~\citep{fitstime},
\item the coordinate formats, following the IAU resolutions~\citep{iau},
\item the N-dimensional maps, with regular or sparse axis that can contain either bins or points and that handle physical units,
\item the metadata, with standard keywords associated to e.g. the instrument, maintainer, data release identifier, data format version, VO standards\footnote{~e.g. \href{https://www.ivoa.net/documents/RM/20070302/index.html}{https://www.ivoa.net/documents/RM/20070302/index.html}}; each of the following data levels should contain metadata information,
\item the provenance information, following the IVOA data model\footnote{~\href{https://www.ivoa.net/documents/ProvenanceDM/}{https://www.ivoa.net/documents/ProvenanceDM/}}; each of the following data levels should contain provenance information.
\end{itemize}

\item[Science-Ready Data.] They are data ready for delivery to users that contain sets of selected events (typically a mix of signal and irreducible background) with a single final set of reconstructed parameters per event. This data level also contains the IRF of the instrument, currently factored into four independent components: the Effective Area, the Energy Dispersion, the Point Spread Function and the Background model. These contain also auxiliary data describing the astronomical, environmental, and instrumental conditions required for science analysis, such as the stable observation interval, pointing position(s), livetime. And this level should contain index tables such that these data can follow the FAIR standards, in particular their findability and access.

\item[Binned Science Data.] These data are produced by binning the spatial, temporal, and/or spectral components of all the Science-Ready Data over a target-specific or a region of interest of the celestial sphere with the user choice of binning. These are in instrumental units, e.g. counts.

\item[Advanced Science Data.] They consist on the astrophysical products derived from the Binned-Science Data (combining the binned events with the binned IRFs), such as spectral energy distributions, light curves, sky maps, and phasograms. These are in physical units such as fluxes, energy or angles.

\item[High-Level Catalog Data.] They contain a collection of astrophysical objects or VHE sources with a description of their properties and associations, such as source components, upper-limits on size or flux, observation periods  or science alerts.
\end{description}

In addition to these formats, the VODF initiative aims to provide tools to check the compliance of a data set with a given format version. These open tools, which use Python and the community-standard {\tt astropy} library, could be used by observatories during their data release process and included in their  Verification and Validation (V\&V) plan.

The definition of such format will be preceded by a data modeling analysis. The data models of each data level could be considered to be released in addition to the data formats. 

The definition and serialisation of sky models are a major bottle neck in the VHE and HE domain. Up to know, there is no convergence between the communities of experiments or analysis software to create standards on models with a their spectral, spatial and/or temporal representation (see for example the difference between \href{https://docs.astropy.org/en/stable/modeling/index.html}{{\tt astropy.modeling}}, the \href{https://cxc.cfa.harvard.edu/sherpa/models/}{Sherpa models} that include the \href{https://heasarc.gsfc.nasa.gov/xanadu/xspec/manual/Models.html}{XSpec ones}, or the stand-alone library \href{https://astromodels.readthedocs.io/en/latest/}{{\tt astromodels}}). Given the complexity of the task, the VODF initiative has postponed to work on this aspect for the moment.

\section{Organisation of the VODF initiative}
\label{sec:org}

THe VODF initiative was formed at the beginning of 2023 and is officially supported by eleven VHE experiments: ASTRI, CTAO, FACT, Fermi-LAT, HAWC, H.E.S.S., IceCube, KM3NeT, MAGIC, SWGO and VERITAS (in alphabetical order). The data format used by the Fermi-LAT experiment was used as a base for the original GADF and has been settled since a long time. Thus, the participation of Fermi-LAT in this initiative will bring in valuable expertise and skills, as well as ensure that the VODF format can serve both ground-based and satellite experiments.

With the official delegates of each experiment, a governance document has been written and it describes the functionning of the initiative. The latter possesses a simple and clear organisation:
\begin{itemize}[noitemsep,nolistsep] 
\item a Steering Committee composed by one official delegate per experiment,
\item three Lead Editors, one per experimental technique (IACT, WCD, Neutrino),
\item and two Conveners.
\end{itemize}
The initiative is fully open and a community-supported project. Any contribution will be considered and evaluated. The initiative is committed to fostering an inclusive community with an established Code of Conduct.\\

The setting and the improvement of the formats will be driven by open contributions from the community (users and experiments). Experts of the VHE domain and of other astronomical fields are welcome to participate, as well as users. All inputs on features, documentation or tools will be discussed. The lead editors will drive the discussions and validate the proposals accepted by consensus. Ultimately, the steering commitee might arbitrate some discussion. The lead editors are responsible for the maintenance of the supporting tools and report the advance to the steering committee. Major orientation plans or roadmaps and major improvements will be prepared by the lead editors and evaluated by the steering committee.\\

The initiative are using several tools to communicate, exchange and develop the standards. A central web site\footnote{~\href{https://vodf.readthedocs.io/en/latest/index.html}{https://vodf.readthedocs.io/en/latest/index.html}} hosted by {\tt readthedocs.org} has been settled and is still under construction. A dedicated exchange workspace\footnote{~\href{vodf-workspace.slack.com}{vodf-workspace.slack.com}} is created on the communication platform {\tt Slack} to exchange between contributors. And a VODF {\tt GitHub} project\footnote{~\href{https://github.com/vodf}{https://github.com/vodf}} hosts the standards and the format documentation, the interface to report changes ({\tt Issues}) and to propose improvements ({\tt Pull requests}), the V\&V tools and the web site files.

\section{Conclusion and perspectives}
\label{sec:pers}

VODF is a newly-born open initiative aiming to create standards for VHE data produced by ground-based gamma-ray and neutrino instruments. Supported officially by eleven astroparticle projects, the VODF initiative aims to settle a new data format for the high-level products of VHE instruments that respects the FAIR principles and follows as closely as possible the IVOA standards. It will propose VHE standards for Common Data Structures, Science-Ready Data, Binned Science Data, Advanced Science Data and High-Level Catalog Data.

The initiative has just been created and standards have not yet been settled. Some data modelling is under way and evaluations of some technical choices are currently made to propose machine-readable formats and to serialise data under this format (e.g. within FITS, YAML, ASDF). The first version of the VODF format will probably be an extension of the GADF format including the experience from CTA and HAWC members. The in-depth data models produced by the Computing Department of CTAO are key elements to take into account. The next versions will  respect the FAIR principles as soon as possible, such that the next open data will be easily findable and accessible. The specifities of the neutrino detectors will be included following the experiences gathered from IRF generation for KM3NeT in a common CTA and KM3NeT analysis using Gammapy. And the IVOA recommandations will be added as much as possible, with the willing of providing feedback to the IVOA. The exact roadmap will be soon settled by the lead editors.

The data formatting is a hidden pillar of Open Science. It permits that open data can follow the FAIR principles and that open software can follow the FAIR4RS principles~\citep{fair4rs}. The VODF initiative is thus an important activity that supports the current and under-construction VHE experiments, and the VHE Science Analysis Tools. It will allow a real interoperability between the different VHE data in order to realise joint multi-wavelength and multi-messenger analyses.


%
%
%

\end{document}